\documentclass[twocolumn,preprintnumbers,amsmath,amssymb,superscriptaddress,subeqn,prl]{revtex4-1}

\usepackage{graphicx}
\usepackage{times}
\usepackage{soul}
\usepackage{color} 
\usepackage{physics}
\usepackage{mathtools}

\def\ket#1{\mathinner{|{#1}\rangle}}

\newlength{\singlecolumn}
\newcommand{\mb}{\mathbf}
\newcommand{\mr}{\mathrm}
\newcommand{\xtalvol}{\ensuremath{\mathcal{V}_\mathrm{xtal}}}

\newcommand{\Ve}{\ensuremath{V^\mathrm{e}}}
\newcommand{\Vh}{\ensuremath{V^\mathrm{h}}}
\newcommand{\VSB}{\ensuremath{V_{S\rightarrow B}}}
\newcommand{\Vif}{\VSB}

\makeatletter
\renewcommand*{\@fnsymbol}[1]{\ensuremath{\ifcase#1\or \dagger\or *\or  \ddagger\or
\mathsection\or \mathparagraph\or \|\or **\or \dagger\dagger
\or \ddagger\ddagger \else\@ctrerr\fi}}
\makeatother

\begin{document}

\title{Origins of singlet fission in solid pentacene from an \textit{ab initio} Green's-function approach}

\author{Sivan Refaely-Abramson}
\thanks{These two authors contributed equally.}
\affiliation{Department of Physics, University of California Berkeley, Berkeley, CA 94720 , USA}
\affiliation{Molecular Foundry, Lawrence Berkeley National Laboratory, Berkeley, CA 94720, USA}

\author{Felipe H. da Jornada}
\thanks{These two authors contributed equally.}
\affiliation{Department of Physics, University of California Berkeley, Berkeley, CA 94720 , USA}
\affiliation{Materials Sciences Division, Lawrence Berkeley National Laboratory, Berkeley, CA 94720, USA}

\author{Steven G. Louie}
\thanks{Corresponding authors: sglouie@berkeley.edu and jbneaton@lbl.gov}
\affiliation{Department of Physics, University of California Berkeley, Berkeley, CA 94720 , USA}
\affiliation{Materials Sciences Division, Lawrence Berkeley National Laboratory, Berkeley, CA 94720, USA}

\author{Jeffrey B. Neaton}
\thanks{Corresponding authors: sglouie@berkeley.edu and jbneaton@lbl.gov}
\affiliation{Department of Physics, University of California Berkeley, Berkeley, CA 94720 , USA}
\affiliation{Molecular Foundry, Lawrence Berkeley National Laboratory, Berkeley, CA 94720, USA}

\begin{abstract}
We develop a new first-principles approach to predict and understand rates of singlet fission with an \textit{ab initio} Green's-function formalism based on many-body perturbation theory. Starting with singlet and triplet excitons computed from a GW plus Bethe-Salpeter equation approach, we calculate the exciton--bi-exciton coupling to lowest order in the Coulomb interaction, assuming a final state consisting of two non-interacting spin-correlated triplets with finite center-of-mass momentum. For crystalline pentacene, symmetries dictate that the only purely Coulombic fission decay from a bright singlet state requires a final state consisting of two inequivalent nearly degenerate triplets of nonzero, equal and opposite, center-of-mass momenta. For such a process, we predict a singlet lifetime of 30 to 70~fs, in very good agreement with experimental data, indicating that this process can dominate singlet fission in crystalline pentacene. Our approach is general and provides a framework for predicting and understanding multiexciton interactions in solids.
\end{abstract}

\maketitle

Harnessing multiexciton generation processes, by which multiple charge carriers may ultimately result from a single photon, is of significant interest for achieving efficiencies beyond the Shockley-Queisser limit of conventional solar cell devices~\cite{Nozik2002, Rabani2010}. One important multiexciton process is singlet fission (SF),  where one photoexcited spin-singlet exciton is converted into two lower-energy spin-triplet excitons~\cite{Smith2010, Smith2013, Congreve2013, Piland2014}. This process is particularly prominent in some organic semiconducting crystals, where significant electron-hole exchange interactions lead to large singlet-triplet splittings~\cite{Kronik2016}. A well-studied example is solid pentacene, where SF timescales of 70-200~fs have been reported based on transient-absorption (TA)~\cite{Rao2010, Wilson2011, Wilson2013, Yost2014} and time-resolved two-photon photoemission spectroscopy~\cite{Chan2011} measurements. Despite intense recent research efforts~\cite{Monahan2015}, the nature of this process still lacks consensus.

Because of the short time scales and challenges associated with direct measurements of triplet states, theoretical calculations of SF are particularly important. While such calculations traditionally use wavefunction-based approaches~\cite{Shavitt2009}, which directly treat correlated bi-exciton states, they must also correctly capture the environmental effects and boundary conditions appropriate to the crystalline condensed phase. Although there has been recent progress in applying coupled-cluster methods in the solid state~\cite{Gruneis2015, McClain2017}, it is still highly challenging to perform such calculations for organic crystals. Several prior studies have approximated organic crystals by small finite clusters, such as a dimer of two pentacene molecules, e.g. in Refs.~\cite{Yost2012, Feng2013, Yost2014, Coto2014, Beljonne2013, Berkelbach2013a, Berkelbach2013b}. While these frameworks have provided insight and fit to measured trends in some cases, they result in simplified descriptions of singlet and triplet excited states, leading to incorrect optical excitation energies, oscillator strengths, and selection rules~\cite{Sharifzadeh2013, Monahan2015}; they involve empirical parameters; and, crucially, they are inconclusive on the details of the SF process. Some emphasize the need for an intermediate dark delocalized ``charge-transfer'' excitation~\cite{Smith2010, Greyson2010, Berkelbach2013a, Yost2014, Beljonne2013}; others highlight the importance of nuclear zero-point and thermal motion in facilitating the fission process~\cite{Arago2015, Zimmerman2011, Zimmerman2013}. Berkelbach \textit{et al.}~\cite{Berkelbach2014} and more recently Tempelaar \textit{et al.}~\cite{Tempelaar2017a, *Tempelaar2017b} incorporated both the crystal environment and symmetry, as well as lattice vibration effects, into an approach that can explain recent reports of SF dynamics, with input from \textit{ab initio} many-body perturbation theory (MBPT) calculations. Very recently, Kryjevski \textit{et al.}~\cite{Krijevski2017} used a MBPT approach to evaluate SF rate in carbon nanotubes with periodic boundary conditions. Although these are important advances, a fully predictive theoretical approach without experimental input that clarifies central contributions to SF associated with triplet excitons carrying center-of-mass momentum $\vb{Q}$, a direct consequence of crystalline translational symmetry, is heretofore lacking.

In this letter we develop and demonstrate a new predictive framework to understand and compute exciton fission in crystalline systems using \textit{ab initio} MBPT. We start with singlet and triplet excitons eigenstates computed from the \textit{ab initio} GW-Bethe-Salpeter equation (GW-BSE) approach~\cite{Rohlfing1998, Rohlfing2000}, and then evaluate the exciton--bi-exciton coupling to lowest order in the Coulomb interaction, assuming a final state consisting of two spatially non-interacting but spin-correlated triplets with equal and opposite center-of-mass momentum. We apply this approach to crystalline pentacene, and uncover a new optically active singlet exciton decay channel, one that involves distinct triplet final states with finite center-of-mass momenta, and that had been unrecognized in previous studies. The computed timescale for this newly predicted process is in good agreement with reported measurements for pentacene crystals, indicating that it can play a dominant role in SF for this and other systems.

\begin{figure}[t]
\begin{center}
\includegraphics[width=\singlecolumn]{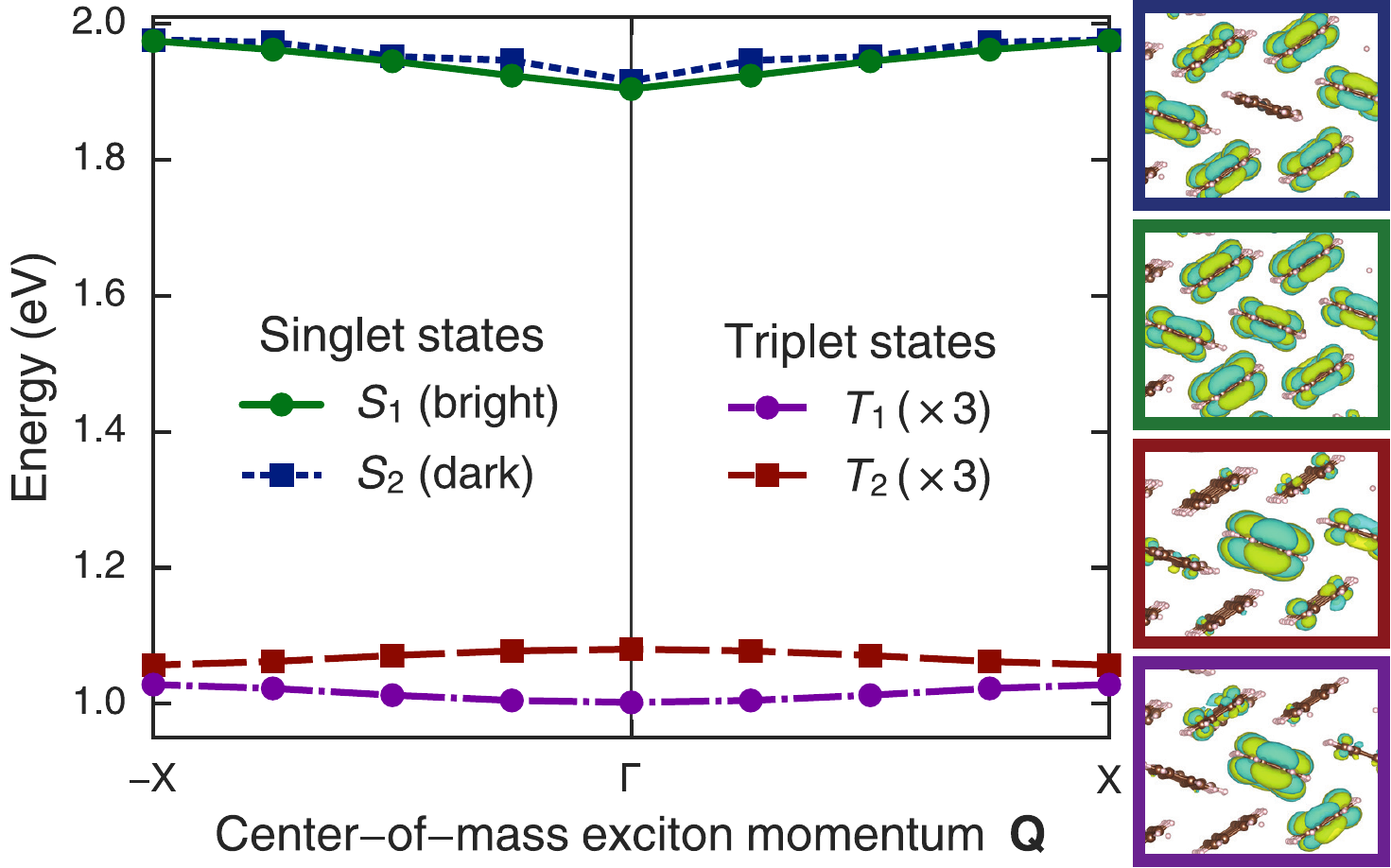}
\centering
\caption{(Color online) Left panel: Excitation energies calculated at different exciton momentum for the two lowest bands of singlet and triplet excitations, where each triplet state is triply degenerate ($m=-1,0,1$). The singlet states are labeled as bright (odd parity) or dark (even parity) at $\vb{Q}=0$. Right panel: exciton wavefunctions for a hole positioned in the middle of the center molecule. The different states correspond to (top to bottom): $S_2$, $S_1$, $T_2$ and $T_1$ at $\vb{Q}=0$.}
\label{fig:Fig1}
\end{center}
\end{figure}

As described below, central to a fast SF process is the presence of two distinct low-energy triplet states -- with energies of about half of the lowest singlet energy -- which disperse into bands as function of $\vb{Q}$. The available phase space of finite-$\vb{Q}$ triplet products is crucial to enable non-zero matrix elements between the triplet products and the initial optically active (or bright) singlets, which are necessarily of odd parity in crystals with inversion symmetry as in the case of solid pentacene.

Our approach to computing the SF decay lifetime $\tau_S$ starts with the standard Fermi's golden rule expression
\begin{equation}
\begin{split}
\tau_{S}^{-1}
	= \frac{2\pi}{\hbar} \frac{1}{\xtalvol} \sum_{B} \left| \Vif \right|^2 
	  	\rho (\Delta E(S{\rightarrow}B)),
\label{eqn:FGR}
\end{split}
\end{equation}
where $S$ is the initial spin-singlet single-exciton state and $B$ a four-particle final bi-exciton state; \xtalvol{} is the crystal volume; \Vif{} is a coupling term; $\rho$ is, most generally, a delta function enforcing the conservation of energy and momentum; and $\Delta E(S{\rightarrow}B) \coloneqq \Omega_{S,\vb{Q}_S} - \Omega_{B,\vb{ Q}_B}$ is the singlet--bi-triplet energy difference, where $\Omega_{S,{\bf Q}_S}$ [$\Omega_{B,\vb{Q}_B}]$ and $\vb{Q}_S [{\bf Q}_B]$ are the energy and center-of-mass momentum of the initial [final] configuration. In this work we consider $\vb{Q}_S=\vb{Q}_B\approx0$, since the initial state is optically excited. Owing to environmental fluctuations, we expect that $\Delta E(S{\rightarrow}B)$ would also be sample dependent.

We focus on the limit for which the bi-triplet pair is nearly non-interacting: as excitons are neutral excitations,  bi-exciton coupling is expected to be significantly weaker than that between the electron and hole that form a single exciton (see SI). In addition, the bi-triplet binding energy is likely much smaller than the broadening $\sigma$ of the initial singlet state. In this limit, we can treat the bi-exciton as a product of two spatially non-interacting but spin-correlated excitons, which we write as
$\ket{B} \approx \ket{T,\vb{Q}; T',\vb{-Q}}$, with excitation energy $\Omega_B \approx \Omega_{T,\vb{Q}} + \Omega_{T',\vb{-Q}}$, where $\ket{B}$ and $\Omega_B$ are the spatial part of the amplitude and the excitation energy associated with the bi-exciton state $B$, respectively; $\ket{T,\vb{Q}}$ and $\Omega_{T,\vb{Q}}$ denote the spatial part of the amplitude and excitation energy of a triplet exciton labeled with a discrete band index $T$ and $\vb{Q}$, which can be obtained from first principles using the GW-BSE approach. Within this assumption, we note that $\Delta E \coloneqq \Delta E(S{\rightarrow}T,T';\vb{Q})=\Omega_{S,0} - \Omega_{T,\vb{Q}} - \Omega_{T',{-}\vb{Q}}$ varies as a function of both the band indices and the center-of-mass momenta of the final triplet states.

The spatial amplitude $\ket{i,\vb{Q}}$ of a singlet or triplet exciton is written as a linear combination of electron-hole quasiparticle transitions,
\begin{equation}
    \label{eqn:T_BSE}
    \ket{i,\vb{Q}} = \sum_{vc\vb{k}} A^{i,\vb{Q}}_{v c \vb{k}} c_{c\vb{k}+\vb{Q}}^\dagger c_{v\vb{k}} \ket{0},
\end{equation}
where $c_{n\vb{k}}$ is a destruction operator for a quasiparticle state at band $n$ and wavevector $\vb{k}$, $\ket{0}$ is the ground state, and $A^{i,\vb{Q}}_{v c \vb{k}}$ are the expansion coefficients obtained from solving the BSE. The quasiparticle transitions entering into the BSE are obtained from the \textit{ab initio} GW approach~\cite{Hybertsen1986}, and hence they already include electronic correlation effects beyond Hartree-Fock. The full bi-exciton state is written as a product of spatial, $\ket{B}$, and spin, $\ket{X}$, components. Since the initial state is a singlet, the final state is restricted to be an overall singlet, and in particular $\ket{X}=\ket{j_1{=}j_2{=}1,J{=}M{=}0}$~\cite{Fetter}. Our framework is formulated appropriately for periodic crystals, where a photogenerated singlet exciton $\ket{S,\vb{Q}_S{=}0}\otimes\ket{j{=}m{=}0}$ can decay to spin-correlated bi-triplet states of the form $\ket{T,\vb{Q}; T',\vb{-Q}}\otimes\ket{X}$~\footnote{See Supplemental Material for further explanation of the evaluation of the interaction spin part}. This is in marked contrast with a dimer (or a finite-cluster) picture of SF, which can, at best, only capture such processes \textit{ad hoc}. 

The quasiparticle wavefunctions and energies are obtained from \textit{ab initio} density functional theory (DFT) and GW calculations, respectively~\footnote{See Supplemental Material for computational details, which includes Refs.~\cite{Siegrist2007,CCDC,PBE1996,espresso,Deslippe2013}.}. Singlet and triplet excitation energies and wavefunctions are obtained by solving the BSE with expansion coefficients $A^{i\mb{Q}}_{vc\mb{k}}$~\cite{Qiu2015}, and shown in Fig.~\ref{fig:Fig1}. All GW-BSE calculations are done using the BerkeleyGW code~\cite{Deslippe2012}. The full wavefunction of the lowest singlet state $S_1$ at zero $\vb{Q}$ is of odd parity and has non-zero oscillator strength for photon absorption, and we refer to this state as being bright; the second-lowest singlet $S_2$ is nearly degenerate but is of even parity and hence possesses zero oscillator strength, and we refer to it as being dark. Our calculated excitation energies of these two low-lying singlet excitons in solid pentacene are 1.9 eV at $\vb{Q}{=}0$, slightly higher than the commonly reported experimental value of 1.83 eV \cite{Duhm2009, Wilson2013}. The singlet dispersion is 0.1~eV, in agreement with previous studies~\cite{Cudazzo2013, Roth2012}. We find two low-lying triplet excitation energies. The energy of the lowest triplet $T_1$ is 1.0~eV at $\vb{Q}{=}0$, slightly higher than the commonly used experimental value of $0.86$~eV~\cite{Burgos1977}, but close to recently reported values~\cite{Ehrler2012}. The second triplet $T_2$ is about 80~meV higher in energy. The triplet states display less dispersion ($\sim30$~meV) than the singlets and, notably, $T_1$ and $T_2$ disperse in opposite directions. From our results, and considering only $\vb{Q}{=}0$, either $S_1$ or $S_2$ could decay into $T_1$ or $T_2$, with an energy difference $\Delta E \approx -0.1$ to $-0.2$~eV. Our calculated $\Delta E$ is slightly negative, unlike the commonly used value for pentacene crystal of slightly positive $\Delta E$; we address this point later in this letter. However, since we use a scattering formalism as opposed to a thermally activated process in this work, our lifetime calculations via Fermi's golden rule are insensitive to the sign of $\Delta E$. 

The difference in energy and bandwidth between the singlet and triplet can be understood from their contrasting characters: the wavefunctions of the triplet excitons in electron-hole relative coordinates are considerably more tightly bound than those of the singlet excitons, in which the wavefunction is distributed over several molecules. This also indicates that the dimer model, and even small clusters, are limited in their ability to accurately capture the nature of low-lying exciton states in crystalline pentacene~\cite{Tiago2003, Sharifzadeh2013, Monahan2015, Rangel2016}. Finally, we note that in the periodic case, both singlet and triplet excitons are delocalized across the crystal due to translation symmetry, which physically gives rise to the center-of-mass quantum number $\vb{Q}$ as a good quantum number for excitons.

\begin{figure}
\begin{center}
\includegraphics[width=\singlecolumn]{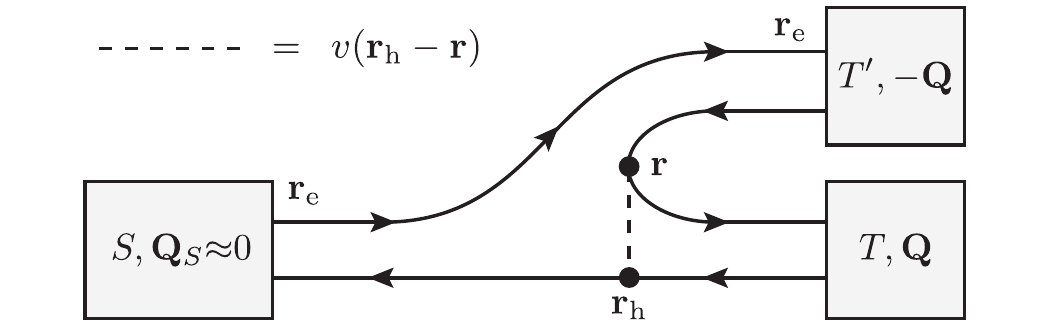}
\centering
\caption{Feynman diagram representing the hole-channel contribution ($\Vh$) to the coupling between one singlet exciton and a pair of triplet excitons (see Eq.~\ref{eqn:meh}). Each gray box represents an incoming or outgoing exciton, solid lines represent quasiparticles that make up each exciton, and the dashed line represent the bare Coulomb interaction.}
\label{fig:Fig2}
\end{center}
\end{figure}

Next, we obtain the coupling term \Vif{} between initial single-exciton and final bi-exciton states. We consider the coupling arising due to electron-electron interactions, although it is straightforward to generalize our approach to include interactions involving phonons as well. Since both the quasiparticle and excitonic states are already dressed by electron-electron interaction, we avoid double-counting the Coulomb interaction by rigorously amputating~\cite{Peskin} the propagators for the initial and final states~\footnote{See Supplemental Material for further explanation of the use of the bare Coulomb interaction, including Ref.~\cite{Benedict2002}}. To lowest order in the Coulomb interaction, \Vif{} can be written as two contributing channels, where a hole channel (electron channel) arises from the scattering of a hole (electron) in the initial singlet exciton, as shown diagrammatically in Fig.~\ref{fig:Fig2} for the hole channel. The overall coupling term is given by
\begin{align}
    \label{eqn:Vif}
\Vif{}
    = \sqrt{\frac{3\xtalvol}{2}} \Big[
      &\Ve(T,T',\vb{Q}) + \Ve(T',T,-\vb{Q})  \\
    - &\Vh(T,T',\vb{Q}) - \Vh(T',T,-\vb{Q}) \Big] \notag,
\end{align}
where \Ve{} and \Vh{} have simple expressions in real space,
\begin{align}
\label{eqn:meh}
\begin{aligned}[t]
V^\mathrm{e/h}(T,T',\vb{Q}) = \int
\dd[3]{{r}}\dd[3]{{r}_\mr{e}}\dd[3]{{r}_\mr{h}}
{{\Psi^*}^{S,\vb{0}}}(\vb{r}_\mr{e},\vb{r}_\mr{h}) \\
{}\times v(\vb{r}_\mr{e/h} - \vb{r})
 \Psi^{T',\vb{-Q}}(\vb{r}_\mr{e},\vb{r}) \,
 \Psi^{T,\vb{Q}}(\vb{r},\vb{r}_\mr{h}),
\end{aligned}
\end{align}
and where $\Psi^{i,\vb{Q}}(\vb{r}_\mr{e},\vb{r}_\mr{h})=\sum_{v c \vb{k}} A^{i,\vb{Q}}_{v c \vb{k}}\psi_{c\vb{k}+\vb{Q}}(\vb{r}_\mr{e})\psi_{v\vb{k}}^*(\vb{r}_\mr{h})$ is the real-space exciton amplitude at the electron/hole coordinate $\vb{r}_\mr{e}$/$\vb{r}_\mr{h}$, $\psi_{n\vb{k}}$ is a quasiparticle wavefunciton, $v$/$c$ labels occupied/unoccupied states, and $v(\vb{r})$ is the bare Coulomb interaction.

Even though we are explicitly treating a many-electron environment, we note that the exciton--bi-exciton interaction relevant to our study rigorously involves the bare, and not screened Coulomb potential. The use of a screened Coulomb interaction would be adding extra exchange-like electron-hole correlation between specific quasi-electrons and quasi-holes of different excitons (see SI). This interaction is already present in an interacting bi-exciton, and should be amputated~\cite{Peskin}. In our current framework, although the final state is not an interacting bi-exciton, the small bi-exciton binding energy estimated from other calculations~\cite{Aryanpour2015} is an indication that either this inter-exciton exchange is not important, or that it should be cancelled by other diagrams of similar magnitude. Indeed, we observe that $|\Vif(\vb{Q}{=}0)|^2$ reduces by only $40\%$ in solid pentacene if we screen the Coulomb interaction in Eq.~\ref{eqn:meh}, suggesting that the final bi-triplet state is weakly correlated in this system.

\begin{figure}
\begin{center}
\includegraphics[width=\singlecolumn]{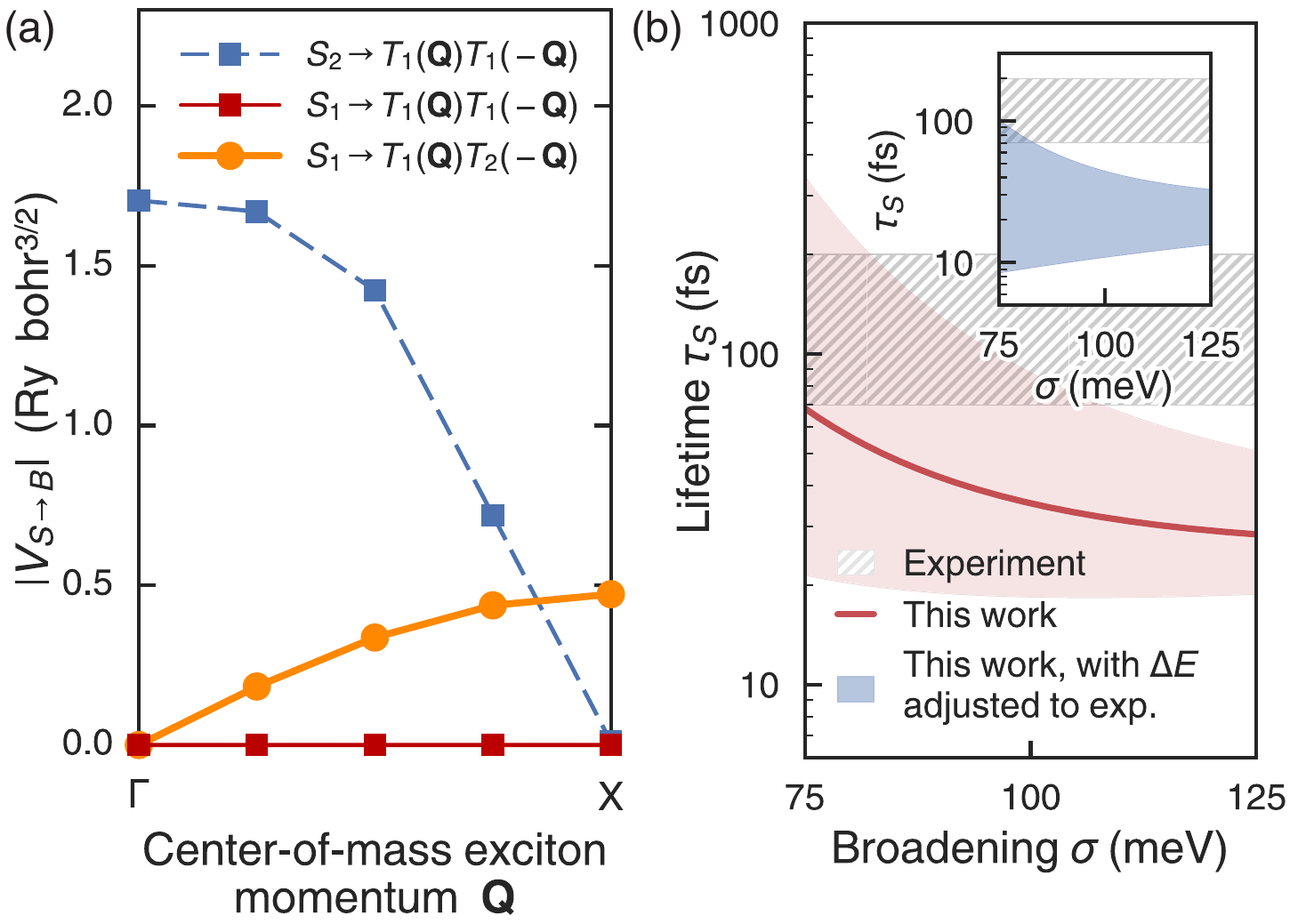}
\centering
\caption{(Color online) (a) SF matrix elements $\Vif{}$ from the lowest bright and dark singlet states (with $\vb{Q}_S{=}0$) to low-lying triplet pairs as a function of $\mb{Q}$, the center-of-mass momenta of an individual triplet state within the pair $\ket{T(\vb{Q})T'(-\vb{Q})}$. Decay from the bright singlet to a pair of non-equivalent triplets ($T{\ne}T'$) via the Coulomb interaction is allowed when $\vb{Q}{\ne}0$ (orange line). (b) Lifetime $\tau_{S}$ of the lowest bright singlet as a function of the broadening parameter $\sigma$. The red line represents SF lifetime obtained directly from our calculation; the red shaded area accounts for uncertainty in $\Delta E(S{\rightarrow}B)$ of up to 50~meV. The inset shows calculated SF lifetimes when excitation energies are shifted in our calculations according to experiment~\cite{Chan2011, Wilson2011, Wilson2013, Yost2014, Bakulin2016}, as discussed in the text.}
\label{fig:Fig3}
\end{center}
\end{figure}

Equations~\ref{eqn:Vif} and \ref{eqn:meh} form the main result of our approach and, together with Eq.~\ref{eqn:FGR}, enable calculations of SF rates to two non-interacting spin-correlated triplets arising from Coulomb interactions for an arbitrary crystal. We use Eq.~\ref{eqn:Vif} to calculate the coupling term between an initial singlet exciton $S$ to all possible low-energy bi-triplet states in solid pentacene of the form $\ket{B}=\ket{T(\vb{Q})T'(-\vb{Q})}$. We consider the case where the final bi-triplet state is a product of two triplet states with the same ($T{=}T'{=}T_1$ or $T{=}T'{=}T_2$) or different ($T'{=}T_1$ and $T'{=}T_2$) exciton band labels. Figure~\ref{fig:Fig3}~(a) shows the resulting coupling matrix elements for an initial bright ($S_1$) or dark ($S_2$) exciton as a function of $\vb{Q}$. The decay channel from a bright singlet to a bi-triplet with the same triplet band label, $\ket{S_1} \rightarrow \ket{T_1(\vb{Q})T_1(-\vb{Q})}$ (red squares), is forbidden and has a strictly zero coupling. The coupling between an initial dark singlet ($S_2$) and any bi-triplet state is allowed (blue squares). But most importantly, a direct decay channel from a bright singlet to a bi-triplet composed of states with distinct band labels, $\ket{S_1} \rightarrow \ket{T_1(\vb{Q})T_2(-\vb{Q})}$ (orange circles), is also allowed for nonzero $\vb{Q}$.

Since the pentacene crystal possesses inversion symmetry, the Hamiltonian commutes with the parity operator $\hat{P}$, and each exciton wavefunction $\Psi$, which contains both the orbital and envelope contributions, can be classified as even or odd; as mentioned, $S_1$ is odd and bright, and $S_2$ is even and dark. $T_1$ and $T_2$ are both odd, and the bi-exciton states comprised of two triplet states with the same exciton band but any center-of-mass momenta, $\ket{T(\vb{Q})T(-\vb{Q})}$, are even. We emphasize that these symmetry conditions are not enforced in dimer models, which, as pointed out in Ref.~\cite{Berkelbach2014}, do not always possess inversion symmetry.

Thus, we obtain an important selection rule for SF in crystals with inversion symmetry: Coulomb matrix elements between an initial bright singlet and a final bi-triplet of the form $\ket{T(\vb{Q})T'(-\vb{Q})}$ can be nonzero if both (i) $\vb{Q}{\neq}0$ and (ii) the two triplets are nonidentical (i.e. $T{\neq}T'$). The identification of the nature and strength of this direct decay process for SF -- one involving direct decay of a bright singlet and originating from inversion and translational symmetry -- is a major result of this work; although a direct Coulomb process has been suggested before for pentacene (see e.g. \cite{Berkelbach2014}), our work suggests that the presence of two symmetry-inequivalent molecules in the centrosymmetric unit cell of crystalline pentacene, and hence two low-energy triplet states with nonzero center-of-mass momenta, is critical for the extremely fast SF decay in this material. If (i) and (ii) are not both satisfied for a system with inversion symmetry, a purely Coulombic decay channel for a bright singlet state is not allowed. In an actual experiment, however, this condition might be partially relaxed, as disorder and ionic vibration can locally break the inversion symmetry.

Having identified the nature of a direct Coulomb channel, we now evaluate the lifetime of the bright $S_1$ exciton for pentacene. Because the two low-energy triplets $T_1$ and $T_2$ display a nearly opposite dispersion, i.e., $\Omega_{T_1}(\vb{Q}) + \Omega_{T_2}(-\vb{Q}) \approx \mathrm{const}$, the SF rates depend sensitively on the SF energy difference $\Delta E$ and the density of final states $\rho(\Delta E)$ in Eq.~\ref{eqn:FGR}. For an extended system, we expect the density of final states to have a finite broadening due to both homogeneous and inhomogeneous sources, which can be estimated, for instance, from the width of the singlet peak in experimental optical spectra. In our work, we replace the delta function with a single-parameter Gaussian, $\rho(\Delta E)=\frac{1}{\sqrt{2\pi\sigma^2}} e^{-(\Delta E)^2/2\sigma^2}$, for a range of typical broadening $\sigma$ of $75$ to $125$~meV~\cite{Duhm2009, Hinderhofer2007}. As shown in Fig.~\ref{fig:Fig3}~(b), this range of $\sigma$ leads to SF decay times of 20 to 70~fs, in good agreement with the reported experimental decay times of 70 to 200~fs~\cite{Chan2011, Wilson2011, Wilson2013, Yost2014, Bakulin2016}. The SF rates obtained here for solid pentacene are robust with respect to the typical uncertainty in the calculated exciton excitation energies. Since we compute the lifetime associated with a scattering process using Fermi's golden rule, small variations in $\Delta E(S_1{\rightarrow}T_1,T_2;\vb{Q}{=}0)\approx-0.16$~eV do not affect the calculated rate much. To illustrate that, we also report rates computed when (1) accounting for an uncertainty in $\Delta E$, i.e., we let $\Delta E \rightarrow \Delta E \pm 50$~meV; and (2) rigidly shifting our computed $\vb{Q}$-dependent singlet and triplet excitation energies to match commonly-reported experiment values, $\Omega_{S_1}(\vb{Q}{=}0){=}1.83$~eV and $\Omega_{T_1}(\vb{Q}{=}0){=}0.86$ to $0.96$~eV~\cite{Burgos1977,Duhm2009,Wilson2013,Ehrler2012}, shown in the inset in Fig.~\ref{fig:Fig3}~(b), so that $\Delta E(S_1{\rightarrow}T_1,T_2;\vb{Q}{=}0)$ varies from $-0.17$ to $0.03$~eV. These additional calculations yield lifetimes of 10 to 300~fs -- the same order of magnitude as reported experimentally and as computed using our GW-BSE excitation energies. Taken together, these calculations support the notion that the Coulomb channel we have identified here for singlet fission is likely an important -- if not the most important -- mechanism for SF in crystalline pentacene.

Although we have focused here on singlet decay due to electron-electron interactions, we stress that other channels, such as those involving phonons, and also bi-exciton fusion~\cite{Frankevich1978, Burdett2013, Bayliss2014}, will also contribute to the effective decay rate, and can in principle be computed using the same framework developed here. Additionally, the nuclear motion is indirectly captured by our model through the homogeneous broadening of the singlet state. However, we expect a direct exciton-phonon channel, and even polaronic effects, to be more important when $|\Delta E|$ is large, in which case phonons are essential for the overall energy conservation in the SF process. In fact, this is precisely the case in the tetracene crystal~\cite{Thorsmolle2009, Grumstrup2010, Smith2013, Arias2016, Yost2014}. The fact that the measured singlet lifetime is at least an order of magnitude longer in these systems is an indication that the underlying mechanism for SF in these systems is indeed different than for pentacene. 

Finally, we make some remarks about the basis set used to describe optical excitations. While the \textit{ab initio} BSE approach gives accurate excitation energies and optical absorption spectra of materials, it does not give the full exciton wavefunction, as it does not directly contain information about multi-particle excitations. However, this knowledge is not necessary to compute SF in crystal pentacene. Due to the timescale of excitation probes used in experiments ($\tau_p = 10$ to 20~fs~\cite{Chan2011, Yost2014, Bakulin2016}), it is appropriate to separate the processes of exciton formation in the sudden approximation and the SF, as we propose here, since we find that $\tau_\mr{SF}>\tau_p$. This separation is also likely valid when crystals are illuminated under sunlight due the low coherence time ($\tau_\mathrm{sun}\sim\frac{h}{4k_B T}\sim 2$~fs)~\cite{Donges1998}.

Our Green's function-based framework clarifies how the electron-electron interaction can be a central mechanism for ultrafast SF in solid pentacene. Our calculations show that this decay takes places at timescales of the order of $\sim50$~fs, without transitions to virtual exciton states or direct phonon-assisted processes.  Finally, our analysis reveals the important role of symmetry in the SF process, and shows that not only energy conservation, but inversion symmetry, the number of symmetry-inequivalent molecules in a unit cell, and the parity of each finite-momentum triplet state play a central role in determining the SF rates in crystals, suggesting future experiments may tune SF rate by altering crystal structure and symmetry.

This work was supported by the Center for Computational Study of Excited-State Phenomena in Energy Materials (C2SEPEM) at the Lawrence Berkeley National Laboratory, which is funded by the U.S. Department of Energy, Office of Science, Basic Energy Sciences, Materials Sciences and Engineering Division under Contract No. DE-AC02-05CH11231, as part of the Computational Materials Sciences Program. Work performed at the Molecular Foundry was also supported by the
Office of Science, Office of Basic Energy Sciences, of the U.S. Department of Energy under the same contract number. S.R.A acknowledges Rothschild and Fulbright fellowships.
We thank Eran Rabani and Naomi Ginsberg for valuable discussions, and Diana Qiu for helpful code development. This research used resources of the National Energy Research Scientific Computing Center (NERSC). 

\bibliography{SFbib}

\end{document}